\documentclass[aps,reprint,twocolumn,pre,10pt,nolongbibliography]{revtex4-2}
\usepackage{amsmath,amsfonts}
\usepackage{graphicx}
\usepackage{amssymb}
\usepackage{enumerate}
\usepackage{mathrsfs}
\usepackage{dcolumn}
\usepackage{textcomp}
\usepackage[colorlinks,citecolor=blue,linkcolor=blue]{hyperref}
\usepackage{natbib}
\usepackage{etoolbox}

\newcommand{\thbar}{\tilde{\hbar}}
\newcommand{\erfc}{\text{Erfc}}
\newcommand{\erf}{\text{Erf}}

\begin{document}
\title{Spacing ratios in mixed-type systems}
\author{Hua Yan}
\email{yanhua@ustc.edu.cn}
\affiliation{CAMTP - Center for Applied Mathematics and Theoretical Physics,    University of Maribor, Mladinska 3, SI-2000 Maribor, Slovenia}
\date{\today} 
\begin{abstract}
  The distribution of the consecutive level-spacing ratio is now widely used as a tool to distinguish integrable from chaotic quantum spectra, mostly due to its avoiding of the numerical spectral unfolding. Similar to the use of the Rosenzweig-Porter approach to obtain the Berry-Robnik distribution of level-spacings in mixed-type systems, in this work we extend this approach to derive analytically the distribution of spacing ratios, for random matrices comprised of independent integrable blocks and chaotic blocks. We have numerically confirmed this analytical result using random matrix theory in paradigmatic models such as the quantum kicked rotor and the Hénon-Heiles system.
\end{abstract}  
\maketitle

\section{Introduction}

A central result in quantum chaos is the marked difference in spectral statistics between integrable and chaotic systems \cite{percival1973regular,haake2013quantum,berry2017half}. The widely accepted Berry-Tabor conjecture \cite{berry1977level} demonstrates that integrable systems follow Poisson statistics of uncorrelated random variables. In contrast, the Bohigas-Giannoni-Schmit (BGS) conjecture \cite{casati1980connection,bohigas1984characterization} asserts that systems with a chaotic semiclassical limit are well described by Random Matrix Theory (RMT), falling into one of the three classical RMT ensembles based solely on the symmetries of the system. These ensembles consist of Hermitian random matrices with independently distributed entries: real for the Gaussian orthogonal ensemble (GOE), complex for the Gaussian unitary ensemble (GUE), and quaternionic for the Gaussian symplectic ensemble (GSE), indexed by different Dyson index $\beta$ \cite{mehta2004random,akemann2011oxford,tao2012topics}. This conjecture has received strong theoretical support \cite{berry1985semiclassical,sieber2001correlations,muller2009periodic}. 

In spectral statistics, the distribution of level spacings between consecutive energy levels is a crucial indicator of spectral correlations and quantum chaos. In integrable systems, levels exhibit clustering, characterized by the Poisson distribution of spacings, while in chaotic systems, RMT statistics predicts level repulsion, with a distinctive power-law behavior in the spacing distribution for small spacings. The power exponent, given by the Dyson index $\beta$, depends solely on the underlying symmetry of the system \cite{brody1973statistical}. In mixed-type systems, where classically regular and chaotic dynamics coexist, the spectral statistics in the semiclassical limit could be a mixing of the Poisson distribution and RMT statistics. In particular, the spacing distribution follows the Berry-Robnik statistics \cite{berry1984semiclassical,prosen1993energy,prosen1994semiclassical,prosen1998berry,prosen1999intermediate}, derived from the Rosenzweig-Porter approach \cite{rosenzweig1960repulsion}, assuming ignorable quantum tunneling between the regular and chaotic regions.  The Hermitian matrix of mixed-type system is block-diagonal, with the chaotic part coming from an ensemble of random matrices and the regular part represented by a diagonal matrix with uncorrelated random variables. The concept of mixed-type systems can be extended to quantum many-body systems lacking a clear classical correspondence, where Hilbert space fragmentation  \cite{moudgalya2022quantum,moudgalya2022hilbert,adler2024observation} causes the Hamiltonian matrix to decompose into dynamically disconnected Krylov subspaces that exhibit distinct dynamics.

Analyzing spacing distributions requires unfolding \cite{porter1956fluctuations}, a procedure transforming the original levels $\varepsilon_j$  to $\bar{\varepsilon}_j = N(\varepsilon_j)$, where  $N(\varepsilon_j)$ is the mean number of levels below $\varepsilon_j$. The unfolded spectrum has unit mean spacing, enabling uniform comparison of fluctuations among spectrums, particularly with random matrix ensembles. Unfolding is straightforward in systems where an analytical form for $N(\varepsilon_j)$ is available, such as in quantum billiards where Weyl's law provides an approximation \cite{ivrii2016100,kac1966can,giraud2010hearing,grebenkov2013geometrical}, or when sufficient large statistics allows for a stable polynomial fit of  $N(\varepsilon_j)$. However, in the absence of an analytical form or sufficient statistics for a reliable fit, unfolding becomes nontrivial. To circumvent the need for unfolding, Ref. \cite{atas2013distribution} introduced the ratios of consecutive spacings, which inherently eliminate the local dependence of spacing distributions on the level density. This approach has also been used in studies of dissipative quantum chaos, from complex spacings to complex spacing ratios \cite{sa2020complex}. 

Furthermore, Giraud \emph{et al.} \cite{giraud2022probing} recently generalized the Rosenzweig-Porter approach to analyze spacing ratios in systems with independent symmetry subspaces, thereby probing potentially hidden symmetries. In this paper, we apply the same generalization to the analysis of spacing ratios in mixed-type systems, thereby providing a more comprehensive description of spacing ratios that goes beyond the purely integrable and chaotic cases. Analytical results from the Wigner surmise (for the chaotic region) and Poisson statistics (for the regular region) show excellent agreement with results from random matrix ensembles.

The paper is structured as follows. In Sec. \ref{sec2}, we review the Rosenzweig-Porter approach for describing Berry-Robnik statistics of level spacings in mixed-type systems and subsequently present analytical results for spacing ratios obtained using the extended approach, as well as a comparison with numerical results from the random matrix ensembles. In Sec. \ref{sec3} applies these results to two paradigmatic physical models, the quantum kicked rotor and the Hénon-Heiles system. We draw our
conclusions in Sec. \ref{sec4}. In the Appendix we provide a detailed derivation of the analytical results and further details about two physical models studied.

\section{Analytical results}
\label{sec2}
According to the BGS conjecture, generic fully chaotic systems exhibit universal spectral statistics that correspond to the Wigner-Dyson ensembles of random matrices. The exact joint probability distribution for ordered eigenvalues $\{\varepsilon_n\}$ (where $\varepsilon_1<\varepsilon_2<\cdots<\varepsilon_N$) of $N$-dimensional matrices from the Wigner-Dyson ensembles is given by \cite{mehta2004random}
\begin{align}
  \label{eq:jpd-rmt}
  P(\varepsilon_1,\cdots,\varepsilon_N)=C_{N\beta}\prod_{n<m}|\varepsilon_m-\varepsilon_n|^\beta \prod_{n=1}^N e^{-\beta \varepsilon_n^2/2},
\end{align}
where $C_{N\beta}$ is a known normalization constant and the Dyson index $\beta$ is given by $\beta=1,2,4$. For these ensembles, it is well known that the nearest-neighbor level spacing, $\hat{s}_n = \varepsilon_{n+1} - \varepsilon_n$, follows a specific distribution after unfolding, $s = \rho \hat{s}$, where $\rho$ represents the density of eigenvalues. This distribution is well approximated by the Wigner surmise, which can be derived from Eq. \eqref{eq:jpd-rmt} for $2\times 2$ matrices,
\begin{align}
  \label{eq:ps}
p(s) = a_\beta s^\beta e^{-b_\beta s^2},
\end{align}
with constants $a_\beta$ and $b_\beta$ determined from the normalization conditions 
\begin{align}
  \int_0^\infty p(s)ds=1,\quad \int_0^\infty sp(s)ds=1,  
\end{align}
depending on the symmetry class of the ensemble. For multidimensional integrable systems, the levels are uncorrelated and form a Poisson process, with a level spacing distribution of $p(s) = e^{-s}$.

The spacing ratio is defined as the ratio between two consecutive spacings, $\hat{s} = \varepsilon_{n+1} - \varepsilon_n$ and $\hat{t} = \varepsilon_{n+2} - \varepsilon_{n+1}$:
\begin{align}
  r = \min\left(\frac{\hat{s}}{\hat{t}}, \frac{\hat{t}}{\hat{s}}\right) = \min\left(\frac{s}{t}, \frac{t}{s}\right), 
\end{align}
where $t = \rho \hat{t}$. This indicates that the spacing ratios for the spectra remain the same before and after unfolding, providing an advantage for using spacing ratios in spectral statistics by avoiding the need for unfolding \cite{oganesyan2007localization}. Similarly, to obtain the Wigner surmise for level spacings, the joint distribution of two consecutive spacings is well approximated by $3\times 3$ matrices, derived from Eq. \eqref{eq:jpd-rmt} as follows:
\begin{align}
  \label{eq:joint-chaotic}
  p(s,t)= A_\beta s^\beta t^\beta (s+t)^\beta e^{-B_\beta(s^2+st+t^2)},
\end{align}
where $A_\beta$ and $B_\beta$ are normalization constant from an integration of Eq. \eqref{eq:jpd-rmt}. Thus, one can derive the spacing ratio distribution of fully chaotic systems \cite{atas2013distribution}
\begin{align}
  p(r)&=\int_0^\infty ds dt \ p(s,t)\delta (r-\min(
    \frac{s}{t},\frac{t}{s}))\nonumber\\
  &=2\int_0^\infty  p(s,rs)sds = \frac{1}{Z_\beta}\frac{(r+r^2)^\beta}{(1+r+r^2)^{1+\frac{3\beta}{2}}},
\end{align}
with $Z_\beta$ the normalization constant. Generally speaking, $p(r)$ has the functional property $p(\frac{1}{r})=r^2p(r)$. For integrable systems, given the Poisson process of energy levels, the joint distribution of consecutive spacings is
\begin{align}
  \label{eq:joint-regular}
  p(s,t)= e^{-(s+t)}.
\end{align}
This results in a spacing ratio distribution $p(r)=\frac{2}{(1+r)^2}$.

In the following, based on the analytical results of the spacing distribution $p(s)$ and the joint distribution of consecutive spacings $p(s, t)$ for both integrable and fully chaotic systems, we review Berry-Robnik statistics for the spacing distribution in mixed-type systems by introducing the Rosenzweig-Porter approach. We then extend this approach to spacing ratios, obtaining an analytical expression for mixed-type systems.

\begin{figure}
  \includegraphics[width=0.6\linewidth]{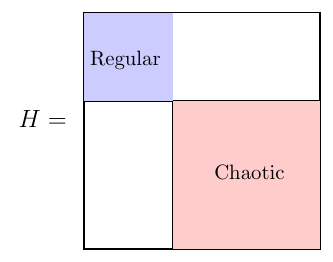}
  \caption{A scheme for the blocked Hamiltonian matrix of mixed systems in the semiclassical limit, illustrated for the simplest case where $ m = 2 $.}
  \label{fig:BR0}
\end{figure}

\subsection{Review of the Rosenzweig-Porter approach for Berry-Robnik statistics}
\label{sec2.1}
\begin{figure}
  \includegraphics[width=0.4\linewidth]{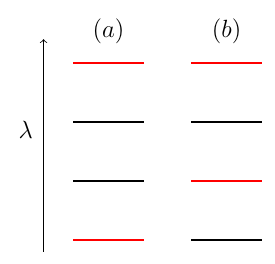}
  \caption{Two configurations for the nearest-neighbor level spacings, from two consecutive level in the center, different colors correspond to spectra of distinct block.}
  \label{fig:BR1}
\end{figure}

Consider a mixed compact phase space, which amounts to an $N$-dimensional Hamiltonian matrix decomposed into $m$ independent blocks in the semiclassical limit, with some blocks being integrable and some  chaotic. The ordered set of energy levels $\{\varepsilon_n\}$, can be characterized by the distribution of nearest-neighbor level spacings. 
For the simplest case where $ m=2 $, one block is regular (integrable) while the other is chaotic, as illustrated in Fig. \ref{fig:BR0}. The spacings are associated with two configurations, as shown in Fig. \ref{fig:BR1}, where different colors correspond to the spectra of each distinct block:
  \begin{enumerate}[(a).]
    \item Configuration for an empty interval $]\varepsilon_n^{(i)}, \varepsilon_{n+1}^{(i)}[$  from $\varepsilon_{n'}^{(j)}<\varepsilon_n^{(i)}<\varepsilon_{n+1}^{(i)}<\varepsilon_{n'+1}^{(j)}$, with $i,j=1, \cdots, m$ labeling different blocks.
    \item Configuration for an empty interval $]\varepsilon_n^{(i)}, \varepsilon_{n'}^{(j)}[$ from $\varepsilon_{n'-1}^{(j)}<\varepsilon_{n}^{(i)}<\varepsilon_{n'}^{(j)}<\varepsilon_{n+1}^{(i)}$.
\end{enumerate}

Given $p(\rho_i \hat{s})$ as the spacing distribution of each spectrum rescaled by its mean level spacing (a process commonly referred to as unfolding), where $\rho_i$ denotes the local density, the resulting superposition of different spectra yields a density $\rho=\sum_i\rho_i$. We introduce two \emph{gap probability} functions:
\begin{subequations}
  \begin{gather}
    f(s)=\int_0^\infty da \ p(s+a), \\
    g(s)=\int_0^\infty da\int_0^\infty db \ p(s+a+b),
  \end{gather}
\end{subequations}
and with the unfolding $s=\rho \hat{s}$, and $\int_0^\infty sp(s)ds=1$, and the related identities are
\begin{align}
  \partial_s f=-p, \quad \partial_s g=-f.
\end{align}

\begin{figure}
  \includegraphics[width=0.65\linewidth]{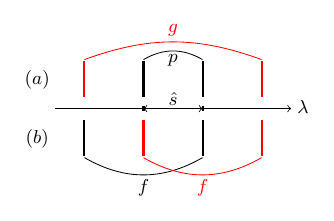}
  \caption{A scheme of spacing distribution for different configurations, corresponding to Fig. \ref{fig:BR1}.}
  \label{fig:BR2}
\end{figure}

In Fig. \ref{fig:BR2}, a scheme for the probability  distribution of spacings is presented, offering a graphical representation of these two functions: $f(s)$ denotes the probability to have a spacing between the (unfolded) nearest-neighbor levels that is greater than or equal to $s$, with one end fixed. $g(s)$ gives the probability of having a spacing greater than or equal to $s$, without any restrictions on either end. The Rosenzweig-Porter approach assumes the block independence and an identical spacing distribution with density $\rho_i$ in each block, from which we get the probability of two configurations
\begin{enumerate}[(a).]
  \item  $p_a(s)=\mu_i^2p(\rho_i \hat{s})g(\rho_j \hat{s})=g(\mu_j s)\partial_s^2 g(\mu_i s)$,
  \item $p_b(s)=\mu_i\mu_jf(\rho_i \hat{s})f(\rho_j \hat{s})=[\partial_sg(\mu_is)][\partial_sg(\mu_js)]$,
\end{enumerate}
where the relative size $\mu_i=\rho_i/\rho$, and $p(\rho_i \hat{s})=p(\mu_i s)$.  The spacing probability $p(s)$ is then the superposition of two configurations summing over $i\ne j$, such that
\begin{align}
  P_m(s)=\sum_{\mathcal{C}\in \{a,b\}}\sum_{i, j}^mp_\mathcal{C}(s)=\partial_s^2 E(s),
\label{eq:br1}
\end{align}
with $\int_0^\infty\ s P_m(s)ds =\int_0^\infty P_m(s)ds=1$, and the gap probability
\begin{align}
  \ E(s)=\prod_{i=1}^{m}g(\mu_i s).
\end{align}

Based on the spacing distribution scheme shown in Fig. \ref{fig:BR2}, one can easily demonstrate that Eq. \eqref{eq:br1} for the expression of $P_m(s)$ is valid for any $m$. This is also known as the Berry-Robnik statistics \cite{berry1984semiclassical}. For systems with time-reversal symmetry, the spacing distribution of chaotic/GOE block is given by Eq. \eqref{eq:ps} with $\beta=1$, namely $p(s)=\frac{\pi}{2}s\exp(-\pi s^2/4)$. Let $\mu_c$ represent the relative size of the chaotic block, the corresponding gap probability is expressed as 
\begin{align}
  \label{eq:mu-c}
  g(\mu_c s)=\erfc(\sqrt{\pi}\mu_c s/2),
\end{align}
and for the integrable spectra $g(s)=e^{-s}$. Therefore, for the simplest case where $m=2$:
\begin{align}
  \label{eq:br2}
  E(s)=e^{(1-\mu_c)s}\erfc(\sqrt{\pi}\mu_c s/2),
\end{align}
while the analytical expression for the spacing distribution $P_m(s)$ is obtained by taking the second-order derivative of $E(s)$ as shown in Eq. \eqref{eq:br2}. Note that Berry-Robnik statistics applies in the semiclassical limit if the Heisenberg time exceeds any classical transport time, according to the principle of uniform semiclassical condensation of Wigner (Husimi) functions of quantum eigenstates \cite{robnik1998topics}. Otherwise, chaotic eigenstates can be quantum localized, with their level spacing distribution approximated by the Brody distribution \cite{prosen1994semiclassical,casati1990scaling2,feingold1991spectral,batistic2010semiempirical,batistic2010semiempirical,batistic2019statistical,batistic2020distribution,wang2020statistical,lozej2021effects,robnik2023recent}.

\subsection{The extended Rosenzweig-Porter approach for spacing ratios}
\label{sec2.2}

\begin{figure}
  \includegraphics[width=1\linewidth]{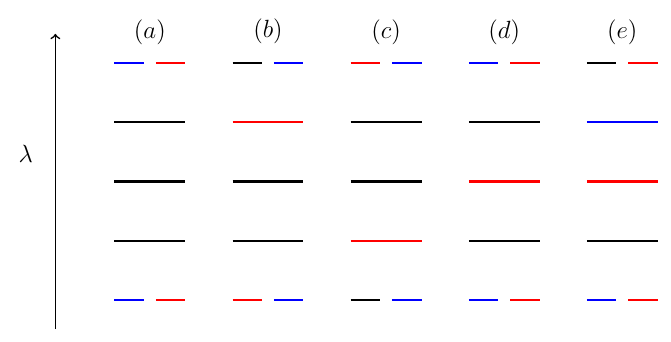}
  \caption{Five configurations for two consecutive spacings, derived from three consecutive levels in the center. The outer levels are at the same height to indicate their relative positions are irrelevant. Different colors correspond to spectra from distinct blocks.}
  \label{fig:SR1}
\end{figure}

Spacing ratios are calculated from two consecutive spacings, which inherently involve three consecutive levels. As a result, in the case of $m=3$ (the $m>3$ cases can be generalized), there exist five configurations corresponding to the patterns of these three consecutive levels, as illustrated in Fig. \ref{fig:SR1}.   Much like the Rosenzweig-Porter approach, which introduces gap probability functions based on the spacing distribution $p(s)$, for the extended Rosenzweig-Porter approach, following Ref. \cite{giraud2022probing}, we introduce two families of \emph{gap probability} functions, based on the distribution of joint consecutive spacings $p(s,t)$. Firstly, there are the one-variable functions:
\begin{subequations}
  \label{eq:gap-func1}
  \begin{gather}
    f(s)=\int^\infty_0da\int^\infty_0db \ p(s+a,b), \\
    g(s)=\int^\infty_0da \int^\infty_0db \int^\infty_0dc \ p(s+a+b,c),
  \end{gather}
\end{subequations}
and then the two-variable functions
\begin{subequations}
  \label{eq:gap-func2}
  \begin{gather}
    e_1(s,t)=\int^\infty_0 da \ p(s+a,t), \\
    e_2(s,t)=\int^\infty_0 da\int^\infty_0 db \ p(s,t+b),\\
    h(s,t)=\int^\infty_0 da\int^\infty_0 db \ p(s+a,t+b).
  \end{gather}
\end{subequations}
These functions are related by the identities 
\begin{align}
  \label{eq:identities}
  \partial_s\partial_t h =p,\ \partial_s h =-e_1,\ \partial_t h=-e_2,\ \partial_s g=-f,
\end{align} 
 with 
 \begin{align}
  \partial_s f = -\hat{p}(s),\quad \hat{p}(s)=\int_0^\infty db\ p(s,b).
 \end{align}

\begin{figure}
  \includegraphics[width=1\linewidth]{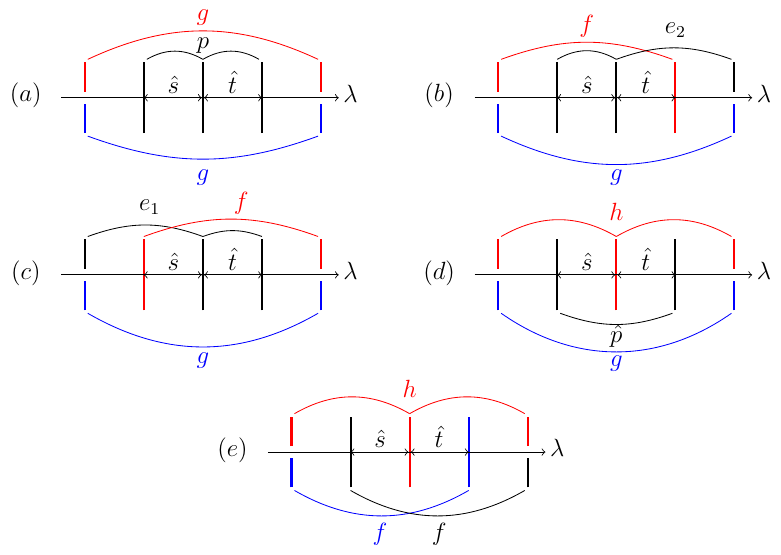}
  \caption{A scheme of the distribution of two consecutive spacings, from which the spacing ratios are calculated, for all the configurations corresponding to Fig. \ref{fig:SR1}.}
  \label{fig:SR2}
\end{figure}

A graphical representation for both families of gap probability functions is shown in Fig. \ref{fig:SR2}, illustrating a scheme for the probability distribution of two consecutive spacings. The marginal probability function $\hat{p}(s)$ differs from the spacing distribution $p(s)$, as visually demonstrated in Fig. \ref{fig:SR2}(d) and Fig. \ref{fig:BR2}. The probability of each configuration can be expressed as a composition of various gap probability functions:
\begin{enumerate}[(a).]
  \item  $p_a(s,t)=\mu_i^3p(\rho_i\hat{s},\rho_i\hat{t})g(\rho_j(\hat{s}+\hat{t}))g(\rho_k(\hat{s}+\hat{t}))\\
 =\mu_i[\partial_s\partial_t h(\mu_i s,\mu_i t)]g(\mu_j(s+t))g(\mu_k(s+t))$,
  \item $p_b(s,t)=\mu_i^2\mu_je_2(\rho_i\hat{s},\rho_i\hat{t})f(\rho_j(\hat{s}+\hat{t}))g(\rho_k(\hat{s}+\hat{t}))\\
  =\mu_i[\partial_s h(\mu_i s,\mu_i t)][\partial_tg(\mu_j(s+t))]g(\mu_k(s+t)),$
  \item $p_c(s,t)=\mu_i^2\mu_je_1(\rho_i\hat{s},\rho_i\hat{t})f(\rho_j(\hat{s}+\hat{t}))g(\rho_k(\hat{s}+\hat{t}))\\
  =\mu_i[\partial_t h(\mu_i s,\mu_i t)][\partial_s g(\mu_j(s+t))]g(\mu_k(s+t)),$
  \item $p_d(s,t)=\mu_i^2\mu_j  h(\rho_j\hat{s},\rho_j\hat{t}) \hat{p}(\rho_i(\hat{s}+\hat{t}))g(\rho_k(\hat{s}+\hat{t}))\\
  =\mu_j h(\mu_j s,\mu_j t)[\partial_s \partial_t g(\mu_i(s+t))]g(\mu_k(s+t)),$
  \item $p_e(s,t)=\mu_i\mu_j  \mu_k h(\rho_j\hat{s},\rho_j\hat{t}) f(\rho_i(\hat{s}+\hat{t}))f(\rho_k(\hat{s}+\hat{t}))\\
  =\mu_j h(\mu_j s,\mu_j t)[\partial_s  g(\mu_i(s+t))][\partial_t g(\mu_k(s+t))].$
\end{enumerate}
The resulting probability is the superposition of all configurations summing over $i\ne j\ne k$, expressed as
\begin{align}
  P_m(s,t)=\sum_{\mathcal{C}\in\{a,b,c,d,e\}}\sum_{i, j, k}^m p_\mathcal{C}(s,t).
\end{align}
Using the identities from Eq. \eqref{eq:identities}, we have
\begin{align}
  \label{eq:sr}
  P_m(s,t)=\partial_s\partial_t E(s,t),
\end{align}
where
\begin{align}
    \label{eq:sr-gap}
  E(s,t) =\sum_i^m \mu_ih(\mu_is,\mu_i t)\prod_{j\ne i}g(\mu_j(s+t)).
\end{align}

By applying the same scheme shown in Fig. \ref{fig:SR2}, the expression for the distribution of two consecutive spacings  in Eq. \eqref{eq:sr} can be generalized to cases where $m > 3$. For the simplest case of $m=2$, this result remains valid, and the proof is provided in Appendix \ref{sec:app1}. One can then obtain the distribution of $r$ as 
\begin{align}
  \label{eq:pr}
  P_m(r)&=\int_0^\infty ds dt \ P_m(s,t)\delta (r-\min(
    \frac{s}{t},\frac{t}{s}))\nonumber\\
  &=2\int_0^\infty ds\ sP_m(s,rs),
\end{align}
since  $P_m(s,t)$ is symmetric in $s$ and $t$, while this symmetry property is inherited from $p(s,t)$. Considering systems with time-reversal symmetry, the joint distribution of consecutive spacings $p(s,t)$ for the chaotic/GOE block is given by Eq. \eqref{eq:joint-chaotic} with Dyson index $\beta=1$, for that 
\begin{align}
  \label{eq:goe-p}
  p(s,t)=\frac{3^7}{32\pi^3}st(s+t)e^{-9(s^2+st+t^2)/(4\pi)}.
\end{align}
As a result, for the chaotic block, two gap probability functions defined in Eq. \eqref{eq:gap-func1}-\eqref{eq:gap-func2} are given by
\begin{align}
  \label{eq:goe-g}
  g(s)=&e^{-9s^2/(4\pi)}-\frac{s}{2}\erfc(\frac{3s}{2\sqrt{\pi}})\nonumber\\
  &-\frac{s}{2}e^{-27s^2/(16\pi)}\erfc(\frac{3s}{4\sqrt{\pi}}),
\end{align}
and 
\begin{align}
  \label{eq:goe-h}
  h(s,t)=&\frac{9(s+t)}{4\pi}V_1(s,t)+\frac{8\pi-27s^2}{16\pi}V_2(s,t)\nonumber \\
  &+\frac{8\pi-27t^2}{16\pi}V_3(s,t),
\end{align}
where 
\begin{equation}
  \begin{aligned}
    V_1(s,t)&= e^{-9(s^2+st+t^2)/(4\pi)}, \\
    V_2(s,t)&=e^{-27s^2/(16\pi)}\erfc(\frac{3(s+2t)}{4\sqrt{\pi}}),\\
    V_3(s,t)&=e^{-27t^2/(16\pi)}\erfc(\frac{3(2s+t)}{4\sqrt{\pi}}).
  \end{aligned}
\end{equation}

For the integrable block, given Eq. \eqref{eq:joint-regular}, the joint distribution yields two simple gap probability functions:
\begin{align}
  g(s)=e^{-s},\quad h(s,t)= e^{-(s+t)}.
\end{align}
Thus, for the simplest case of $m=2$, with $\mu_c$ representing the relative size of the chaotic block, Eq. \eqref{eq:sr-gap} gives
\begin{align}
  \label{eq:m2-ratio}
  E(s,t)=e^{-\mu_0(s+t)}\big[\mu_0g(\mu_c(s+t))
  +\mu_c h(\mu_cs,\mu_ct)\big],
\end{align}
where $\mu_0=1-\mu_c$, is the relative size of the regular parts. In Appendix \ref{sec:app2}, we have derived from $E(s,t)$ the expression of $P(r)$
\begin{align}
  \label{eq:integral1}
  P(r)= \sum_{i,j=0}^4\sum_{k=1}^4\int_0^\infty P^k_{ij} s^k e^{-a_is^2}e^{-bs}\erfc(c_j s)\ ds,
\end{align}
where $b = \mu_0(1+r)$, and $a_i$, $c_j$, and $P^k_{ij}$ are polynomials of both $\mu_c$ and $r$, with $a_0 = c_0 = 0$. These integrals can be obtained from another integral
\begin{align}
  \label{eq:integral2}
  \int_0^\infty dx \ xe^{-ax^2-bx}\erfc(cx).
\end{align}
For all $k \ge 1$, the integral in Eq. $\eqref{eq:integral1}$ can be analyzed by taking derivatives with respect to $b$. Although this integral has a closed-form expression involving Owen's function, but it is too lengthy to be practical. The explicit form and its evaluation are given in the Supplemental Material \cite{suppl}.  For mixed-type systems with multiple chaotic blocks, the lengthy expression of $P_m(r)$ can be derived from Eq. \eqref{eq:sr}-\eqref{eq:pr}, although not in a closed form. Alternatively, an approximation of the error function can be used to simplify the integrals, yielding an approximate closed form, as detailed in Appendix \ref{sec:app2}. 

Notably, the extended Rosenzweig-Porter approach can be easily applied to other ensembles of random matrices, such as GUE and GSE, for modeling chaotic blocks, based on the underlying symmetry of the system.

\subsection{Comparison with numerical results from the random matrix ensembles}

We can now validate the analytical results against numerical data from random matrix ensembles. The regular block is modeled by a diagonal matrix with independent, identically distributed (i.i.d.) Gaussian random diagonal elements. The chaotic block is modeled by a random matrix $H_c$ from GOE, rescaled by its size as $H_c/\sqrt{N_c}$, to make the spectrum of chaotic blocks having the same support in $[-\sqrt{2},\sqrt{2}]$,  because from semicircle law, $H_c$ has support in $ [-\sqrt{2N_c},\sqrt{2N_c}]$. Thus, the relative size $\mu_i = \rho_i/\sum\rho_i =N_i/N$, where $N_i$ is the size of the random matrix. In the $m=2$ case, the relative size of the chaotic block is  $\mu_c=N_c/N$. In all numerical simulations, the size of random matrices is set as $N=1000$, with $1000$ realizations.

\begin{figure*}
  \includegraphics[width=0.8\linewidth]{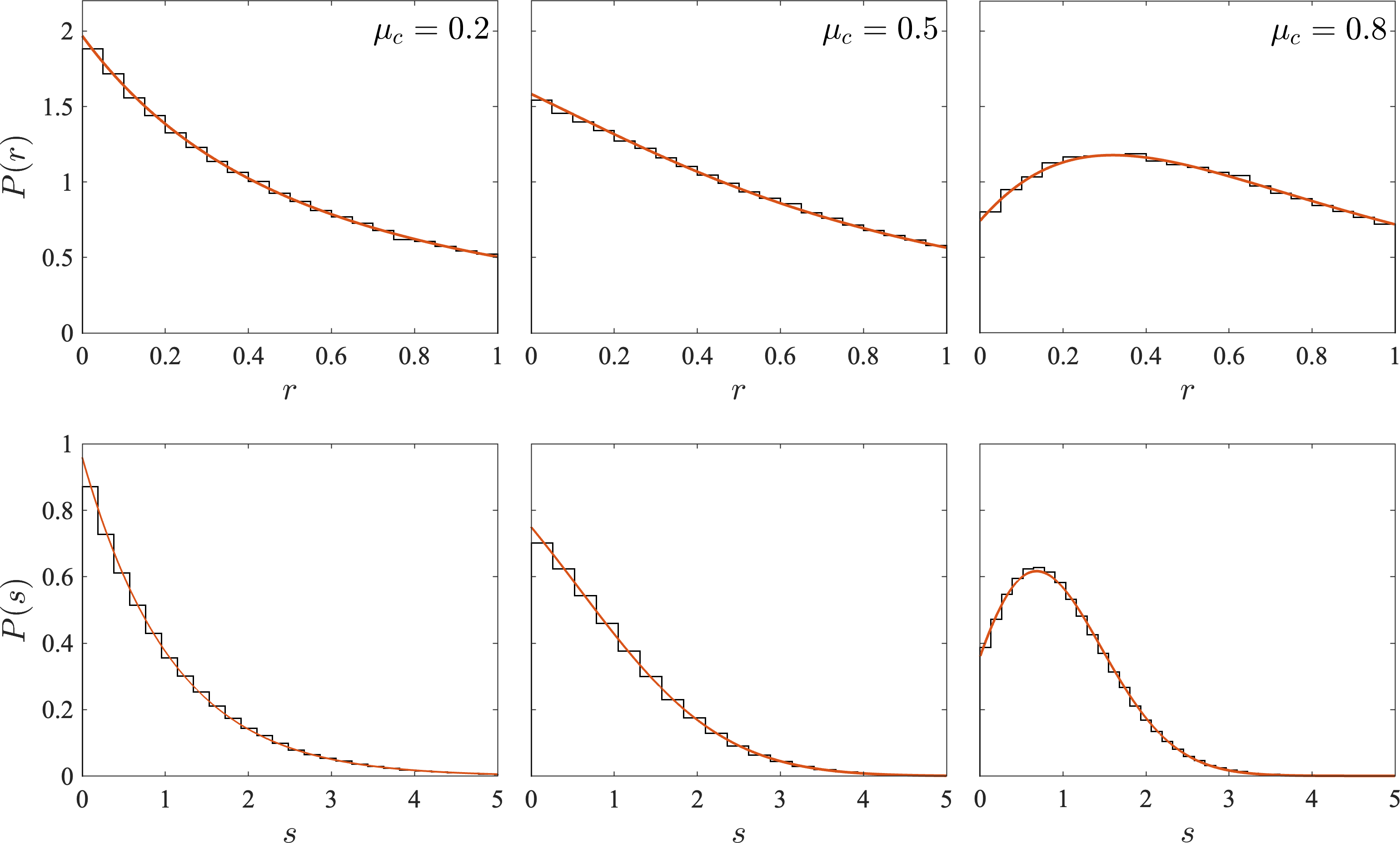}
  \caption{Top panels: From left to right, the histograms of spacing ratios are from random matrices of size $N=1000$ across 1000 realizations for different chaotic block sizes, $\mu_c = N_c/N$, for $m=2$ blocks. The regular block is modeled by a diagonal matrix with i.i.d. Gaussian random elements. Solid lines represent predictions from Eq. \eqref{eq:integral1}. Bottom panels show the corresponding spacing distributions for different $\mu_c$, with solid lines indicating the Berry-Robnik distribution derived from Eq. \eqref{eq:br2}.}
  \label{fig:pr-m2}
\end{figure*}

\begin{figure}
  \includegraphics[width=1\linewidth]{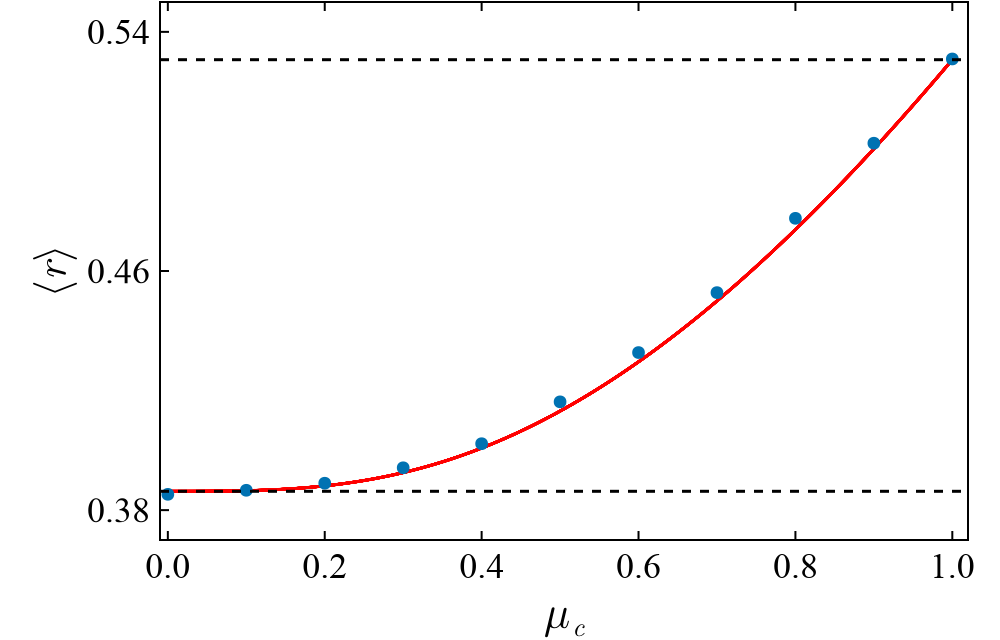}
  \caption{Average spacing ratio $\langle  r \rangle$ as a function of $\mu_c$, the relative size of the chaotic block, for the $m=2$ case. The solid line represents the analytical results, while the points correspond to numerical results obtained from an ensemble of random matrices. Dashed lines represent $\langle r\rangle_P$ for Poisson (bottom) and $\langle r\rangle_{GOE}^N$ for GOE (top).}
  \label{fig:mean-r}
\end{figure}

The mean level spacing ratio, $\langle r \rangle = \int rP(r) \, dr$, exhibits a small deviation from a $3 \times 3$ GOE matrix (implying the Wigner surmise), where $\langle r \rangle_{GOE}^{3} = 4 - 2\sqrt{3} \approx 0.53590$. This contrasts with the numerical result for large $N$ ($N \gg 1$), which converges to $\langle r \rangle_{GOE}^{N} = 0.5307$, as clearly demonstrated in Ref.~\cite{atas2013distribution}. In this work, we use the latter results, and slightly rescale the analytical result of  $\langle r\rangle$ as $ \langle r\rangle_P + a(\langle r\rangle -\langle r\rangle_P)$, where 
\begin{align}
a = \frac{\langle r\rangle_{GOE}^N -\langle r\rangle_P}{\langle r\rangle_{GOE}^3 -\langle r\rangle_P}\approx 0.96524,
\end{align}
and $\langle r\rangle_P=2\ln2-1$ is the value of Poisson ensemble, and we compare this rescaled $\langle r\rangle$ with the numerical results in the following. Fig. \ref{fig:pr-m2} presents both the distribution of spacing ratios and spacings from the random matrix ensemble, along with the analytical results for comparison, in the $m=2$ case, for different values of $\mu_c$. It demonstrates perfect agreement, both for the spacing ratios with Eq. \eqref{eq:integral1} and for the spacings with Berry-Robnik statistics. In Fig. \ref{fig:mean-r}, we compare numerical results of  $\langle r\rangle$ as a function of $\mu_c$, with the analytical prediction. Again, there is good agreement, and it also shows for $\mu_c \lesssim 0.2$, $\langle r\rangle$ remains close to the Poisson ensemble value.

\section{Numerical results from physical models}
\label{sec3}
In this section, we compare the analytical spacing ratios derived from random matrix theory with the numerical results obtained from two paradigmatic models: the quantum kicked top and the Hénon-Heiles system. 

\subsection{Quantum Kicked Rotor}
The Hamiltonian of one-dimensional rotor periodically kicked by a position-dependent amplitude,  reads
\begin{equation}
    \label{eq:kr}
    H = \frac{p^2}{2}+k\cos \theta\sum_n\delta(t-n\tau),
\end{equation}
where $\tau$ and $k$ are period and strength of the kicks, the angular momentum operator $p$ and the angular operator $\theta$ are a pair of canonical variables. 
The Floquet operator is defined as the time-evolution operator for each period
\begin{align}
  U=e^{-ip^2\tau/2}e^{-ik\cos \theta}=e^{-i P^2/2\thbar}e^{-iK\cos \theta/\thbar},
\end{align}
where we rescaled $P=p\tau$, $K=k\tau$, and have taken $\hbar=1$. The dynamics of the system is determined by two parameters: the effective Planck's constant $\thbar=\tau$ and the classical nonlinear parameter $K$. 

The Hilbert space can be represented by the eigenfunctions $\{|n\rangle\}$ of $P$ with $P|n\rangle =n\thbar|n\rangle$, $\langle \theta|n\rangle=\exp(in\theta)/\sqrt{2\pi}$, $n$ being an integer. Then the matrix elements of $U$ are given by
\begin{align}
  U_{n',n}\equiv \langle n'|U|n\rangle=(-i)^{n-n'}J_{n-n'}(K/\thbar)e^{-i n'^2\thbar/2},
\end{align}
where $J_{n-n'}(K/\thbar)$ is a Bessel function of the first kind. In the case $\thbar=2\pi\sigma$, where $\sigma=M/N$ is rational with $M$ and $N$ being coprime integers (also referred as the quantum resonant values \cite{chang1986evolution,izrailev1990simple,kanem2007observation}), the Floquet operator defines a dynamical system on a torus. As a side note, usually $\thbar=4\pi M/N$ is referred as the quantum resonance condition of any order \cite{guarneri2009spectrum}. It is straightforward to verify that for $MN$ being even, we have
\begin{align}
  U_{n' + Nl, n + Nl} = U_{n', n}, \quad (l = 0, \pm 1, \ldots),
\end{align}
which reflects a translational symmetry in momentum space. This can be expressed as $ [U, T_N] = 0 $, where $ T_N $ is the translation operator by $N$ steps, defined by $ T_N |n\rangle = |n + N\rangle $.

This discrete translational symmetry implies that the quasi-energy eigenstates $ |\phi\rangle $ of the Floquet operator $ U $, satisfying $ U|\phi\rangle = e^{-i\epsilon}|\phi\rangle $ where $\epsilon$ is the quasi-energy, take the form of Bloch states:
\begin{align}
  \label{eq:bloch}
  \phi_{s+Nl}\equiv\langle s+Nl|\phi\rangle = e^{-il\alpha}\phi_{s,\alpha}, \quad \alpha \in [0, 2\pi),
\end{align}
where $1 \le s \le N$, and $\alpha$ is the Bloch phase in momentum space. Additionally, $\phi_{s,\alpha}$ is the eigenstate of a reduced $N\times N$ unitary matrix $U_\alpha$, 
\begin{align}
  (U_\alpha)_{s,s'} \equiv \langle s|U_\alpha|s'\rangle =\sum_{l=-\infty}^\infty U_{s,s'+Nl} e^{-il\alpha}.
\end{align}
In the following, we choose $\alpha=0$, implying the periodic boundary condition due to the parity exists in $U$, resulting in the parity in $U_0 \equiv U_\alpha\ (\alpha=0)$. Correspondingly, the quasi-energy eigenstates in phase representation have to be either even or odd $\phi(\theta)=\langle \theta|\phi\rangle =\pm\phi(-\theta)$, which in the momentum representation has the form $\phi_n =\pm \phi_{-n}$, and in the reduced momentum space $ \phi_s =\pm \phi_{-s}=\pm \phi_{N-s}$. One can restrict the study of $U_0$ to the invariant subspace of odd states, and the matrix elements (see details in Appendix \ref{sec:app3}) are
\begin{align}
  \label{eq:torus-parity}
  \langle s|\mathcal{P}_-U_0 &\mathcal{P}_-|s'\rangle 
 =\frac{1}{N}\exp(\frac{-is^2\thbar}{2}) \sum_{j=0}^{N-1}\exp\big(\frac{-iK\cos\frac{2\pi j}{N}}{\thbar}\big)\nonumber \\
 \times &\left[\cos\frac{2\pi(s-s')j}{N}-\cos\frac{2\pi(s+s')j}{N}\right],
\end{align}
where the projection operators are
\begin{align}
  \mathcal{P}_\pm|s\rangle=|s,\pm\rangle =(|s\rangle \pm |N-s\rangle)/\sqrt{2},
\end{align}
 with $1\le s,s' \le N_1-1$ for $N=2N_1$. It should be noted that for the even parity, the basis states $|N/2,+\rangle$ and $|N,+\rangle =|0,+\rangle$ are not normalized.  If $N$ is odd such as $N=2N_1+1$, representation of the matrix elements in the odd subspace is the same as Eq. \eqref{eq:torus-parity} except that $1\le s,s'\le N_1$ \cite{casati1990scaling}.  Numerically, all matrix elements can be efficiently generated using the Fast Fourier Transform, commonly applied in studies of dynamics. The infinite extension of Bloch states in momentum space indicates the absence of localization at resonant values. However, when $N$ is much larger than the system's intrinsic localization length, the eigenfunction distributions at the boundaries of the unit torus space become exponentially small and behave as effectively localized \cite{tian2010theory}.
 \begin{figure}
  \centering
  \includegraphics[width=1\linewidth]{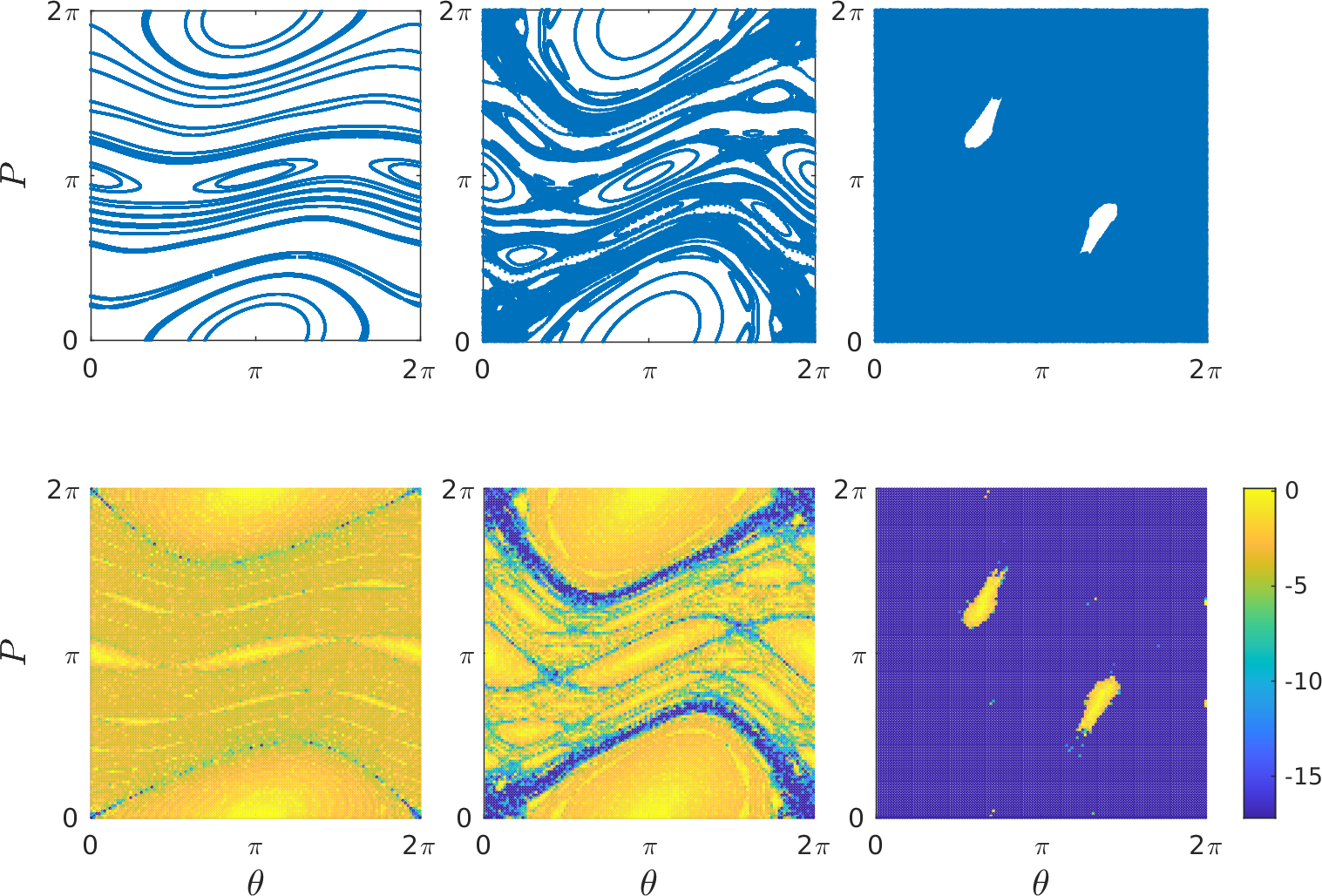}
  \caption{The Poincar\'e section generated from 30 random trajectories of $10^4$ iteration (top panel) and the corresponding logarithmic values of SALI (bottom panel) on the phase space discretized by a grid of size $1000\times 1000$ small square cells, of the same area, after 100 iterations. Regions colored blue correspond to chaotic orbits, the yellowish indicates ordered motion, and the intermediate suggests sticky trajectories. From left to right, the kick strength $K\simeq 0.5, 0.9716, 5$.}
  \label{fig:kr_sali}
\end{figure}

\begin{figure}
  \centering
  \begin{minipage}[b]{\columnwidth}
    \includegraphics[width=0.85\linewidth]{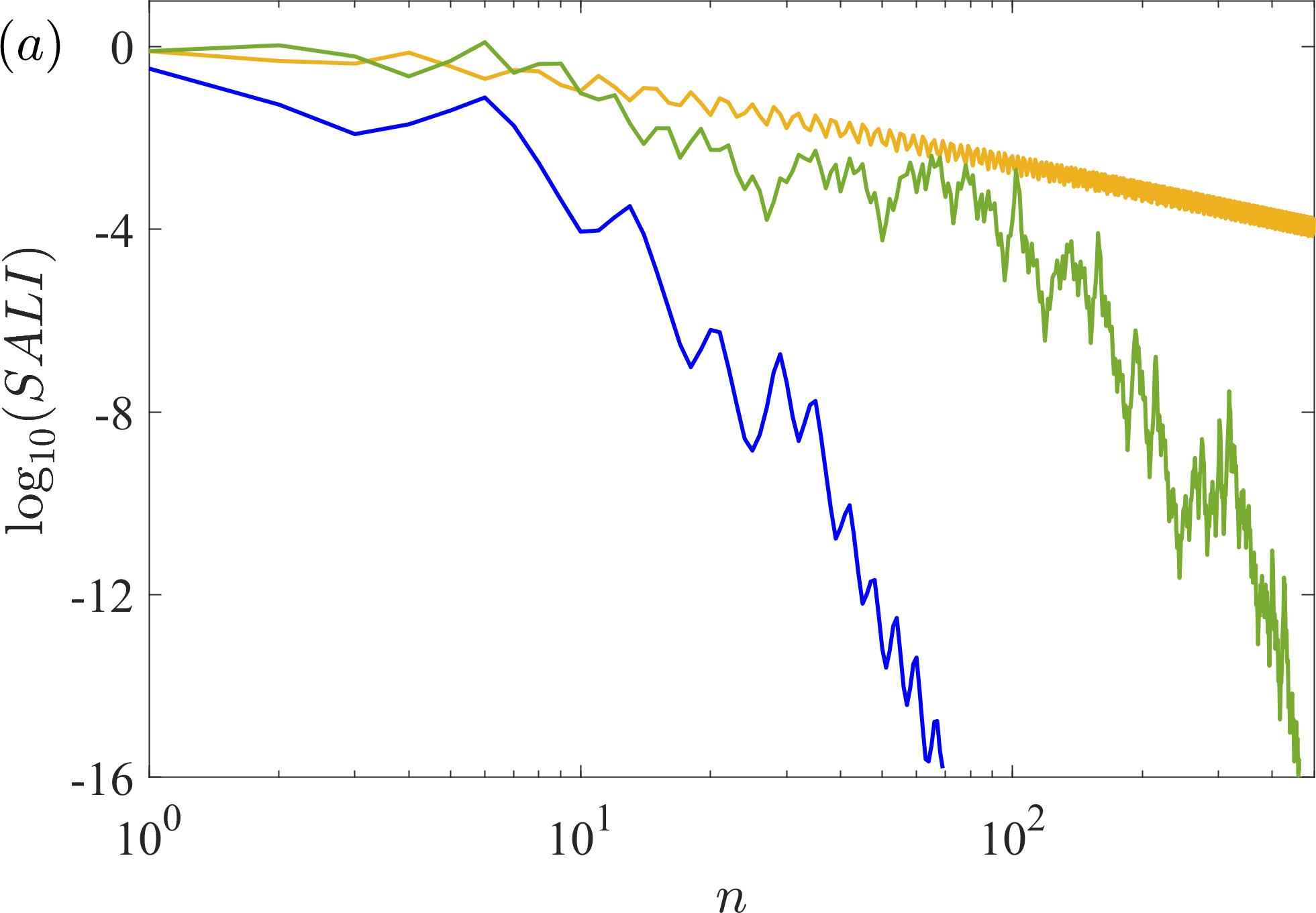}
  \end{minipage}
\begin{minipage}[b]{\columnwidth}
  \includegraphics[width=0.85\linewidth]{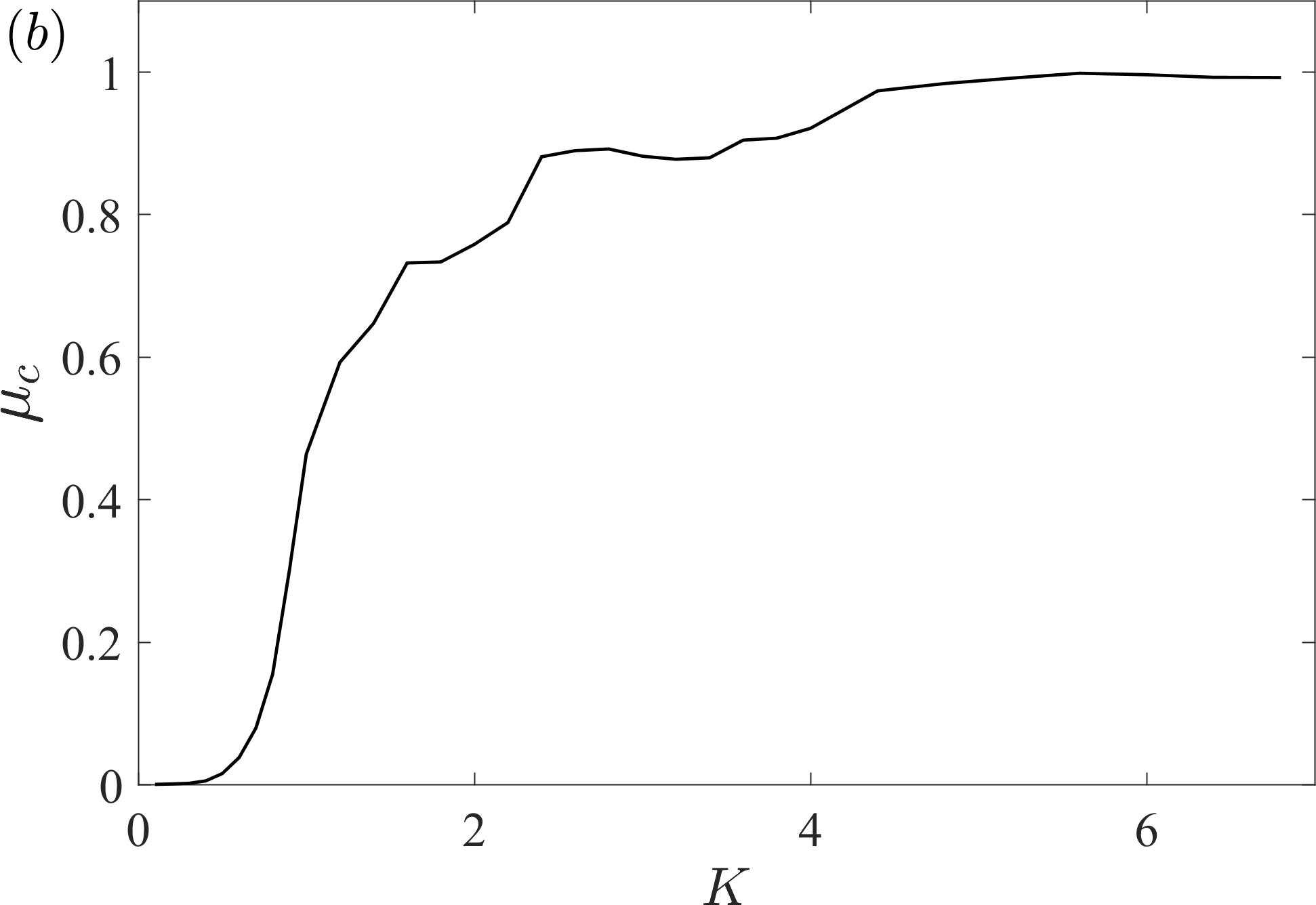}
\end{minipage}
  \caption{(a) Evolution of three typical behaviors of the SALI with respect to numbers of iteration $n$, the initial points are chosen from the phase space of the case with $K_c\approx 0.9716$. (b) The classical chaotic fraction $\mu_c$ in the Chirikov standard map as a function of $K$. The grid size of points is $1000\times 1000$, the calculation of SALI runs 1000 iterations for each point, while the initial points end with $\log_{10}(\text{SALI}) <-8$ are chaotic.}
  \label{fig:kr_sali_detail}
\end{figure}


\begin{figure*}
  \includegraphics[width=0.82\linewidth]{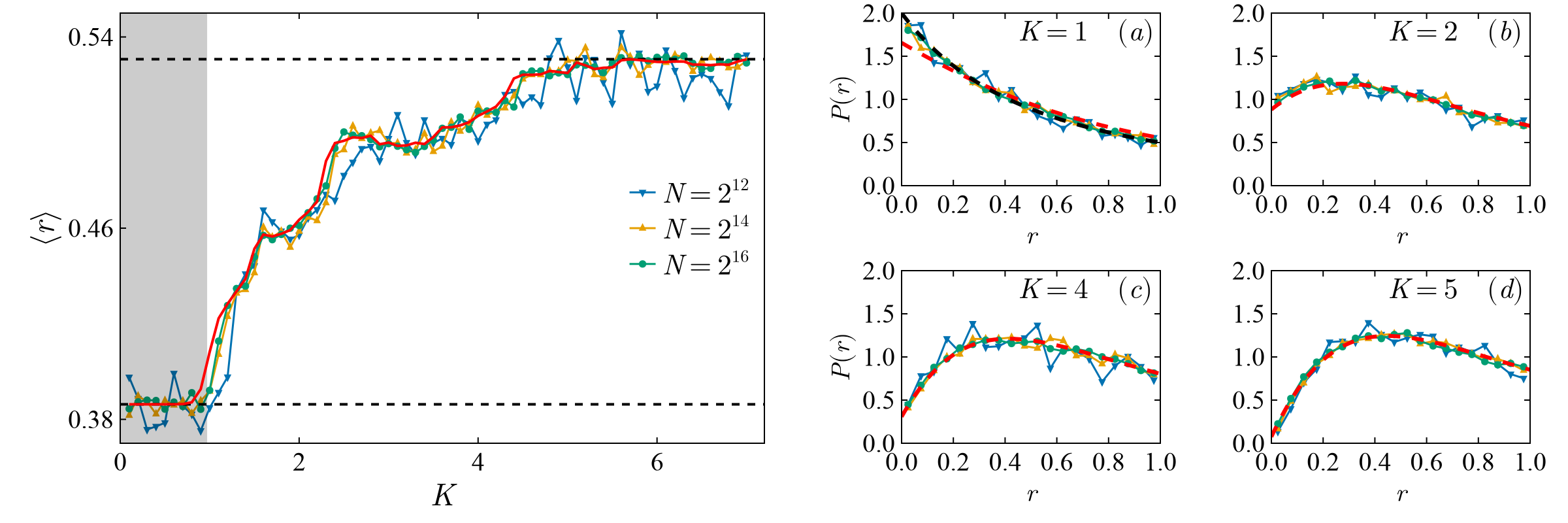} 
  \caption{Left panel: Average spacing ratio $\langle r\rangle$ vs. kicking strength $K$ for varying system sizes $N$ ($\langle r\rangle$ is averaged over two parities, with $\tilde{\hbar} = 2\pi/N$). The solid line shows the $m=2$ analytical result from Eq. \ref{eq:integral1} using $\mu_c$ from Fig. \ref{fig:kr_sali_detail}(b). The two dashed lines are the same as those in Fig. \ref{fig:mean-r}. The gray stripe marks the regime $K \le K_c$, where chaotic regions are disconnected. Right panels (a)-(d) compare $P(r)$ distributions with analytical results (red dashed lines) for different system sizes and $K$. The black dashed line in (a) shows $P(r)$ for the Poisson ensemble.}
\label{fig:qkr-ratio}
 \end{figure*}

 The translational invariance in quantum kicked rotor on resonance can be understood in the classical limit, obtained by taking $\thbar\to 0$ and $k\to \infty$, while keeping $K=kT$ fixed. In this limit, the classical dynamics is described by the Chirikov standard (area-preserving) map \cite{chirikov1979universal}:
 \begin{equation}
   \label{eq:Chirikov}
   \begin{aligned}
     P_{n+1}&=P_n + K\sin \theta_n, \\ \theta_{n+1}&=\theta_n + P_{n+1} \pmod{2\pi}.
   \end{aligned}
 \end{equation}
 This map is invariant under the translation $P \to P+2\pi M$, indicating that in the classical limit, the system is defined on a torus where $P$ is taken modulo $2\pi M$ (in this work, we set $M=1$). The Chirikov standard map $\mathbf{x}_{n+1}=f(\mathbf{x}_n)$ as the classical limit of quantum kicked rotor, is symplectic
\begin{align}
  \frac{\partial f}{\partial \mathbf{x}_n}=\begin{pmatrix}
    1 & K\cos\theta_n \\
  1  & K\cos\theta_n+1  \\
    \end{pmatrix},
\end{align}
where $\mathbf{x}_n=(P_n,\theta_n)$. As an effective indicator of chaos, the smaller alignment index (SALI)  \cite{bountis2012complex} relies on the evolution of deviation vectors $\mathbf{w}$ from a given orbit, given by the corresponding tangent map $\mathbf{w}_{n+1}=(\partial{f}/{\partial \mathbf{x}_n})\cdot \mathbf{w}_n$.  For Hamiltonian system of two degrees of freedom, there exist only one positive Lyapunov exponent $L_1$, and it can be proven that SALI$(t)\propto e^{-L_1t}$. In Fig. \ref{fig:kr_sali} we show the Poincar\'e sections and the logarithmic values of SALI of a dense grid of points of the phase space, plotted using assigned colors accordingly, for three different kick strengths. It verifies that below the critical value $K<K_c$ ($K_c\approx 0.9716$), invariant KAM curves restrict the variation of $P$ (transport in  the $P$ direction) and the phase space features separate chaotic components of significant size, while for $K\gtrsim K_c$, a critical golden KAM curve is broken, transport happens from one part of the chaotic component to the other, but may be impeded by the remaining contori.

The SALI plots offer more details about the transition to chaos, specifically the breakdown of invariant tori and the intricacies of the boundary between the chaotic component and the integrable portion, where varying degrees of stickiness manifest. Shown in Fig. \ref{fig:kr_sali_detail}(a), starting from the top, there are three distinctive orbits: the integrable orbit (because the orbit is restricted on 2D torus, SALI of an invariant tori follows a power law decay), the one exhibiting stickiness, originating from the boundary between the chaotic and integrable regions, and the chaotic orbit characterized by exponential decay of SALI. Therefore,  we would define the classical chaotic fraction, denoted as $\mu_c$, as the relative size of initial conditions for which the value of the SALI falls below a critical threshold after certain times of iteration, effectively distinguishing chaotic orbits from integrable ones:
\begin{align}
  \mu_c=\frac{1}{4\pi^2}\int \chi_c(\theta,P)d\theta dP,
\end{align}
where $\chi_c$ denotes the characteristic function of the chaotic component, which takes the value of 1 on chaotic region and zero otherwise. The chaotic fraction $\mu_c$ measures the transition from integrable dynamics with $\mu_c=0$ to the fully chaotic dynamics $\mu_c=1$.  Here, we use the same notation $\mu_c$ as in Eq.~\eqref{eq:mu-c}, where it denotes the relative size of the chaotic block in the Hamiltonian matrix. For mixed-type systems in the semiclassical limit, this is justified because the classical chaotic fraction equals the relative size of the chaotic quantum spectrum, according to the principle of uniform semiclassical condensation of Wigner functions of quantum eigenstates \cite{robnik1998topics}.

In Fig. \ref{fig:kr_sali_detail}(b), we illustrate how $\mu_c$ varies with the kick strength $K$, resembling the results obtained by the recurrence time approach in Ref. \cite{lozej2020stickiness} for $K \gtrsim K_c$. This indicates that in this regime, only a single chaotic region exists in the classical phase space. The difference occurs  in the regime $K_0 \le K \lesssim K_c$ (where $K_0$ denotes the approximate transition point for $\mu_c > 0$), as multiple disconnected chaotic regions are present in this range. This is clearly reflected in the average spacing ratio, as well as its distribution, as shown in Fig. \ref{fig:qkr-ratio}: for $K > K_c$, the spacing ratios converge toward the analytical results for $m=2$ blocks from Eq. \eqref{eq:integral1}, with increasing system size. For $K$ values close to $K_c$, the $m=2$ result fails, because the chaotic regions are not well-connected, implying $m > 2$. Specifically, at $K=1$, a kick strength slightly larger than $K_c$, where the classical fraction $\mu_c \simeq 0.46$, Fig. \ref{fig:qkr-ratio}(a) shows that the distribution of spacing ratios agrees with the Poissonian result, $P(r) = 2/(1+r)^2$, rather than the $m=2$ result.

\begin{figure*}
  \includegraphics[width=0.85\linewidth]{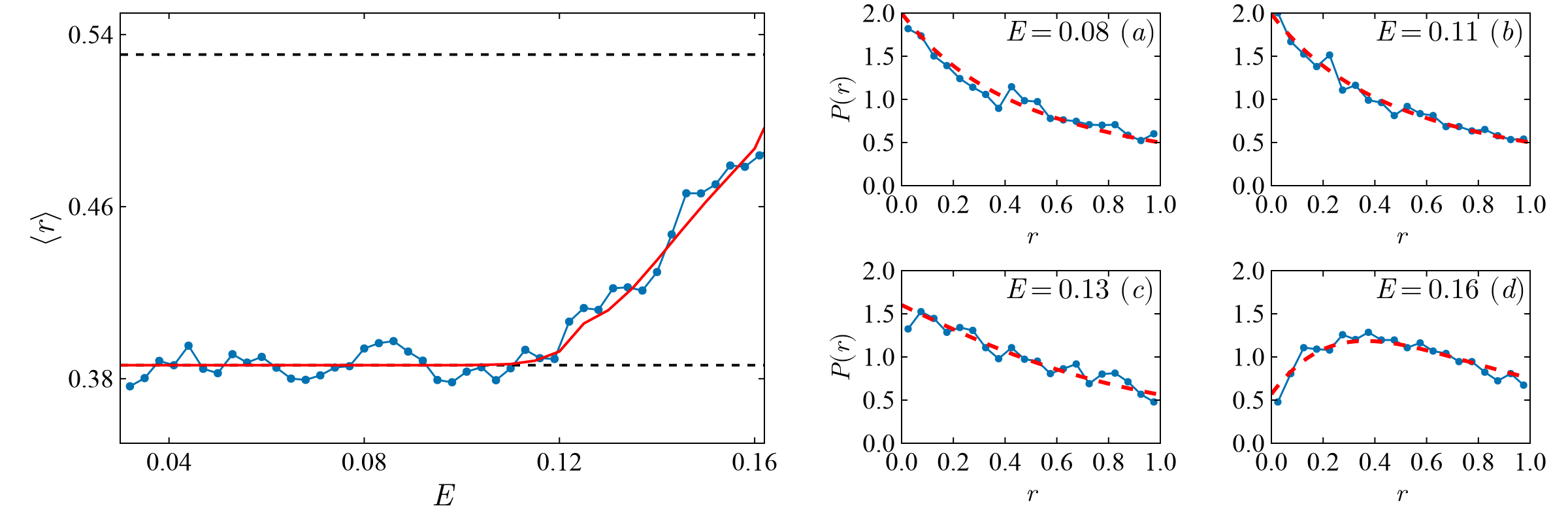} 
  \caption{Left panel: Average spacing ratio $\langle r\rangle$ as a function of energy $E$ for doublets, with the radial quantum number truncated at $N = 1800$, resulting in an irreducible Hilbert space size of $\mathcal{N}_D \sim 5.4 \times 10^5$.  The solid line shows the $m=2$ analytical results using $\mu_c$ from Ref. \cite{yan2024chaos}. The Planck constant is set as $\hbar(E) = (E/E_s)\hbar_c$, where $\hbar_c = 3 \times 10^{-4}$ and $E_s = 1/6$, to effectively calculate 1200 levels within an energy interval approximately $1\%$ of each energy. Right panels (a)-(d) show the distribution of spacing ratios for different energies, with dashed lines representing the analytical results.}
  \label{fig:henon-ratio}
 \end{figure*}

\subsection{H\'enon-Heiles system}
The H\'enon-Heiles Hamiltonian \cite{henon1964applicability}, describing the stellar motion in a 2D galactic potential, is given by
\begin{align}
  H=\frac{1}{2}\sum_{i=1}^2(p_i^2+q_i^2)+\alpha(q_1^2q_2-\frac{1}{3}q_2^3),
\end{align}
with $\alpha$ a parameter governing the anharmonicity (in this work, we set $\alpha=1$). The classical escape energy of the Hénon-Heiles system is given by $ E_s = 1/6\alpha^2 $. Below $ E = 0.5E_s $, which represents the order-chaos threshold energy derived from concavity-convexity analysis, all orbits lie on well-defined invariant tori, indicating that the system is integrable. The SALI method for Hamiltonian dynamics based on the equations of the tangent map obtained by linearizing the difference equations of a symplectic map as the following
\begin{align}
    \dot{\textbf{w}}=\Omega \cdot \nabla^2H(\textbf{x})\cdot \textbf{w}, \quad \Omega = \big[\begin{smallmatrix}
        0 & -I_d  \\
        I_d & 0  \\
        \end{smallmatrix}\big],
\end{align}
where $\textbf{x}=(\textbf{q}, \textbf{p})$ and the deviation vector $\textbf{w}=\delta \textbf{x}$, $\nabla^2 H(\textbf{x})$ is the Hessian matrix, $I_d$ being $d$-dimensional identity matrix ($d=2$). Unlike quantum maps, the H\'enon-Heiles system shows different dynamics across the energy spectrum,  the chaotic fraction $\mu_c$ should be defined as the relative Liouville volume measure of the chaotic part of the phase-space as 
\begin{align}
    \mu_c = \frac{\Phi_c}{\Phi}=\frac{\int d\textbf{q}d\textbf{p}\chi_c(\textbf{q},\textbf{p})\delta(E-H(\textbf{q},\textbf{p}))}{\int d\textbf{q}d\textbf{p}\delta(E-H(\textbf{q},\textbf{p}))},
\end{align}
where $\Phi$ is the phase space volume of the entire energy surface, and $\Phi_c$ the Liouville phase space volume of the chaotic region. In Ref. \cite{yan2024chaos}, We have studied in detail the transition to classical chaos in the H\'enon-Heiles system, particularly in the three-particle Fermi-Pasta-Ulam-Tsingou (FPUT) model, which, with periodic boundary conditions, is equivalent to the Hénon-Heiles system.

For the quantization, because of the $C_{3\nu}$ symmetry in the potential,  we define $\hat{q}_\pm = \hat{q}_1\pm i\hat{q}_2, \
\hat{p}_{\pm}=\hat{p}_1\pm i\hat{p}_2$, and introduce the rotated bosonic operators as
\begin{equation}
\label{eq:rotated-bosonic}
\begin{aligned}
   a_\pm &= \frac{1}{\sqrt{2}}(a_1\mp i a_2)=\frac{1}{2\sqrt{\hbar}}(\hat{q}_\mp +i\hat{p}_\mp) ,\\
   a_\pm^\dagger &= \frac{1}{\sqrt{2}}(a_1^\dagger \pm i a_2^\dagger)= \frac{1}{2\sqrt{\hbar}}(\hat{q}_\pm - i\hat{p}_\pm), 
\end{aligned}
\end{equation}
where the annihilation and creation operator fulfill the canonical commutation relations $[a_i,a_j^\dagger]=\delta_{ij}$, with $i,j\in\{1,2,\pm\}$. Therefore, $\hat{n}$ the number operator and the angular momentum operator $\hat{\ell}=(\hat{q}_1\hat{p}_2-\hat{q}_2\hat{p}_1)/\hbar$ fulfill
\begin{equation}
\begin{aligned}
    &\hat{n}=a_1^\dagger a_1+a_2^\dagger a_2=a_+^\dagger a_++ a_-^\dagger a_-,\\
    &\hat{\ell}= i(a_2^\dagger a_1 -a_1^\dagger a_2)=a_+^\dagger a_+-a_-^\dagger a_-,
\end{aligned} 
\end{equation}
and the quantized Hamiltonian is given as 
\begin{align}
    \hat{H}=\hbar(\hat{n}+1)-i\alpha(\hat{q}_+^3-\hat{q}_-^3)/6.
\end{align}
In circular two-mode basis states $|n,l\rangle$, defined as the simultaneous eigenfunctions of  $\hat{n}$ and $\hat{\ell}$, where $l=-n,-n+2,\cdots,n\  (n\in \mathbb{N}_0)$, $n$ is the radial quantum number and $l$ is the orbital angular momentum (OAM), we have the expressions for matrix elements of the cubic coupling terms, in the same basis. Additionally, in this basis, we can easily categorize the basis states according the irreducible representations of the $C_{3\nu}$ symmetry (see Appendix \ref{sec:app4} and more details in Ref. \cite{yan2024chaos}).  

To obtain spectral statistics over a wide energy range, we need a sufficient number of energy levels, $N(E, \delta E)$, within a narrow interval $[E - \delta E/2, E + \delta E/2]$, where $\delta E = \eta E$ and $\eta \ll 1$ to ensure consistent classical dynamics. We introduce a varying Planck constant, $\hbar(E) = (E/E_s)\hbar_c$, as the semiclassical density of states is roughly linear with $E$. Parameters in Fig. \ref{fig:henon-ratio}, including radial quantum number truncation, were tested to ensure eigenvalue convergence below escape energy $E_s$, with levels below $E_s$ under one-third of the truncated spectrum. Using the chaotic fraction $\mu_c$ measured by the SALI method, Fig. \ref{fig:henon-ratio} compares spacing ratios from the  Lanczos algorithms (based on Krylov subspace) with the $m=2$ analytical result in the quantum Hénon-Heiles system, showing good agreement, for both the average spacing ratio and the distributions.  Figure \ref{fig:mean-r} shows that for $\mu_c \lesssim 0.2$, $\langle r\rangle$ remains close to the Poisson ensemble value. This explains why, for $E \lesssim 0.12$ (at $E=0.115$, $\mu_c\simeq 0.21$), $\langle r\rangle$ fluctuates around the Poisson value, and the integrability-breaking point in quantum spectrum statistics, observed from spacing ratios in Hénon-Heiles, lags behind the classical transition at $0.5E_s = 1/12$. We have carried out the study for a Hilbert space dimension of around $5.4 \times 10^5$, yet fluctuations in the spacing ratio statistics remain. These fluctuations can be reduced by moving deeper into the semiclassical limit, i.e., by increasing the radial quantum number truncation, which presents a challenge for future studies.

\section{Concluding remarks}
\label{sec4}
In this work, we extend the Rosenzweig-Porter approach, originally used to derive the Berry-Robnik distribution for level spacings in mixed-type systems, to analytically derive the distribution of spacing ratios based on the Wigner surmise, avoiding the need for numerical spectral unfolding. This method considers random matrices composed of separate integrable and chaotic blocks, effectively capturing the statistical behavior of systems with both integrable and chaotic regions in their phase space. We verified our analytical results with numerical simulations using random matrix ensembles and models like the quantum kicked rotor and the Hénon-Heiles system. The numerical results closely match the analytical predictions. A more detailed study of spacing ratios in mixed-type systems can be conducted in quantum mushroom billiards, where the classical phase space consists of exactly two regions: one regular and the other chaotic, with the chaotic region being connected \cite{matic2025}. The highly efficient scaling method of Vergini and Saraceno \cite{vergini1995calculation} allows access to a deeper semiclassical limit than is generally achievable in other Hamiltonian systems.

One possible application of this approach is to characterize the relative size of integrable or chaotic regions in mixed-type systems by analyzing quantum spectra, offering insights into the interplay between order and chaos and quantifying their contributions to spectral properties. In this work, we focus on quantum few-body systems with a well-defined classical limit. The approach can also be applied to Hilbert space fragmentation that typically occurs in quantum many-body systems, where the Hamiltonian matrix  consists of several dynamically disconnected Krylov subspaces in a specific basis.

There are two open problems worth addressing. One is the analytical result for spacing ratios, corresponding to the Brody distribution, due to its relevance in quantum localization \cite{batistic2010semiempirical,batistic2010semiempirical,batistic2019statistical,batistic2020distribution,wang2020statistical,lozej2021effects,robnik2023recent}. To avoid localization effects, we extend the system size to  $ 2^{16} $ in our study of the quantum kicked rotor, highlighting the need for a clear theoretical approach to spacing ratios for quantum localization. The other is the generalization of this approach to study complex spacing ratios in dissipative quantum systems, such as the dissipative quantum kicked top, where in the mean-field limit, strange attractors and limit cycles may exist in the classical phase space. 

\section{Acknowledgements}
The author thanks Prof. Marko Robnik for 
the discussions and careful reading of the manuscript. This work was supported by the Slovenian Research and Innovation Agency (ARIS) under the grants J1-4387 and P1-0306.

\appendix
\begin{appendix}

\section{Distribution of two consecutive spacings for \texorpdfstring{$m=2$}{m=2}}
\label{sec:app1}

\begin{figure}[h]
  \includegraphics[width=1\linewidth]{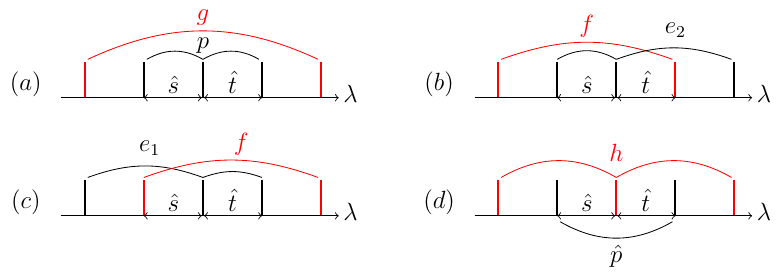}
  \caption{A scheme of the distribution of two consecutive spacings, for the case of $m=2$, where  four configurations exist. Different colors denote spectra from different blocks.}
  \label{fig:sr-simple}
\end{figure}

For the simplest case $m=2$, the distribution of two consecutive spacings is simpler than $m\ge 3$ cases. Fig. \ref{fig:sr-simple} illustrates four different configurations for the distribution of two consecutive spacings, with the gap probability functions defined in Sec. \ref{sec2.2}. The probability of each configuration can be expressed as a composition of various gap probability functions:
\begin{enumerate}[(a).]
  \item  $p_a(s,t)=\mu_i^3p(\rho_i\hat{s},\rho_i\hat{t})g(\rho_j(\hat{s}+\hat{t}))\\
  =\mu_i[\partial_s\partial_t h(\mu_i s,\mu_i t)]g(\mu_j(s+t))$,
  \item $p_b(s,t)=\mu_i^2\mu_je_2(\rho_i\hat{s},\rho_i\hat{t})f(\rho_j(\hat{s}+\hat{t}))\\
  =\mu_i[\partial_s h(\mu_i s,\mu_i t)][\partial_tg(\mu_j(s+t))],$
  \item $p_c(s,t)=\mu_i^2\mu_je_1(\rho_i\hat{s},\rho_i\hat{t})f(\rho_j(\hat{s}+\hat{t}))\\
  =\mu_i[\partial_t h(\mu_i s,\mu_i t)][\partial_s g(\mu_j(s+t))],$
  \item $p_d(s,t)=\mu_i^2\mu_j  h(\rho_j\hat{s},\rho_j\hat{t}) \hat{p}(\rho_i(\hat{s}+\hat{t}))\\
  =\mu_j h(\mu_j s,\mu_j t)[\partial_s \partial_t g(\mu_i(s+t))],$
\end{enumerate}
summing over $i\ne j$ for all configurations, we have
\begin{align}
  \label{eq:sr-simple}
  P_m(s,t)=\sum_{\mathcal{C}\in\{a,b,c,d\}}\sum_{i, j}^m p_\mathcal{C}(s,t)=\partial_s\partial_t E(s,t), 
\end{align}
where $E(s,t)$ is the same as given in Eq. \eqref{eq:sr-gap}, which is
\begin{align}
  E(s,t) =\sum_i^m \mu_ih(\mu_is,\mu_i t)\prod_{j\ne i}g(\mu_j(s+t)).
\end{align}

\begin{widetext}
\section{Closed form of the integral and its approximation}
\label{sec:app2}
Eq. \eqref{eq:m2-ratio} gives the gap probability of the simplest case $m=2$ 
\begin{align}
  E(s,t)=e^{-\mu_0(s+t)}\big[\mu_0 g(\mu_c(s+t))
  +\mu_c h(\mu_cs,\mu_ct)\big],
\end{align}
the joint distribution 
\begin{align}
  P(s,t)&= \partial_s\partial_t E(s,t)=e^{-\mu_0(s+t)}[\mu_0^3g+\mu_0^2\mu_ch+\mu_0\partial_s\partial_tg+\mu_c\partial_s\partial_t h-\mu_0^2\partial_t g-\mu_0\mu_c\partial_th-\mu_0^2\partial_sg-\mu_0\mu_c\partial_sh]\nonumber\\
  &=e^{-\mu_0(s+t)}[\mu_0^3g+\mu_0^2\mu_ch+\mu_0\partial_s\partial_tg+\mu_c\partial_s\partial_t h-2\mu_0^2\partial_t g-\mu_0\mu_c\partial_th-\mu_0\mu_c\partial_sh],\nonumber\\
  &=e^{-\mu_0(s+t)}[\mu_0^3g+\mu_0^2\mu_c (h+2f)+\mu_0\mu_c^2 (\hat{p}+e_1+e_2)+\mu_c^3p],
\end{align}
here $h=h(\mu_cs,\mu_ct)$ and $g=g(\mu_c(s+t))$, and 
\begin{align}
  e_1(s,t)=\frac{243}{32\pi^2} t(2s+t)e^{-9 (s^2+st+t^2)/(4\pi )}+\frac{81 }{128 \pi ^2}t\left(8 \pi -9 t^2\right) e^{-27t^2/(16 \pi)}  \text{Erfc}\left(\frac{3 (2 s+t)}{4 \sqrt{\pi }}\right),
\end{align}
with $e_1(s,t)=e_2(t,s)$. For $f=f(s+t)$, $\hat{p}=\hat{p}(s,t)$, there are
\begin{align}
  f&=\frac{9}{4\pi}(s+t)e^{-9(s+t)^2/(4\pi)}+\frac{1}{2}\text{Erfc}\left(\frac{3(s+t)}{2\sqrt{\pi}}\right)+\frac{8\pi- 27(s+t)^2}{16\pi}e^{-27(s+t)^2/(16\pi)}\text{Erfc}\left(\frac{3(s+t)}{4\sqrt{\pi}}\right), \nonumber\\
\hat{p}&=\frac{243}{32\pi^2}(s+t)^2e^{-9(s+t)^2/(4\pi)}+\frac{81}{128\pi^2}(s+t)(8\pi-9(s+t)^2)e^{-27(s+t)^2/(16\pi)}\text{Erfc}\left(\frac{3(s+t)}{4\sqrt{\pi}}\right).
\end{align}
The distribution of spacing ratios, $P(r)$, can be expressed using the expressions for $g$, $h$, and $p$ from Equations \eqref{eq:goe-p}-\eqref{eq:goe-h} as follows:

\begin{equation}
  \label{eq:integrals}
  P(r) = 2 \int_0^\infty ds \, s P(s, rs) = \sum_{i,j=0}^4 \sum_{k=1}^4 \int_0^\infty P^k_{ij} \, ds \, s^k e^{-a_i s^2} e^{-bs} \text{erfc}(c_j s), \quad k \ge 1,
\end{equation}
where $b = \mu_0(1+r)$, and $a_i$, $c_j$, and $P^k_{ij}$ are polynomials of both $\mu_c$ and $r$, with $a_0 = c_0 = 0$. The explicit form of this expression as well as its evaluation is given the Supplemental Material \cite{suppl}.
These integrals can be obtained from another integral
\begin{align}
  \label{eq:integral}
  \int_0^\infty dx \ xe^{-ax^2-bx}\erfc(cx),
\end{align}
one can deduce for all values of $k\ge 1$ in Eq. \eqref{eq:integrals} by deriving with respect to $b$. The integral in Eq. \eqref{eq:integral} can be rewritten as 
\begin{align}
  \int_0^\infty dx \ xe^{-ax^2-bx}\erfc(cx)&=\int_0^\infty dx \ xe^{-ax^2-bx}-\int_0^\infty dx \ xe^{-ax^2-bx}\erf(cx),
\end{align}
with
\begin{align}
  \label{eq:integral-1}
  \int_0^\infty dx \ xe^{-ax^2-bx}\erf(cx)=&\frac{\sqrt{\pi}b}{4a^{3/2}}e^{\frac{b^2}{4a}}\left[4T(-\frac{bc}{2\sqrt{a(a+c^2)}},-\frac{\sqrt{a}}{c})-1-\erf(-\frac{bc}{2\sqrt{a(a+c^2)}})\right]\nonumber\\
&+\frac{c}{2a\sqrt{a+c^2}}e^{\frac{b^2}{4(a+c^2)}}\erfc(\frac{b}{2\sqrt{a+c^2}}),
\end{align}
where $T(h,p)$ denotes Owen's function, defined as 
  $T(h,p) = \frac{1}{2\pi}\int_0^p \frac{e^{-h^2(1+x^2)/2}}{1+x^2}dx.$
Although the simplest case allows for a closed-form integral for $P(r)$ from Eq. \eqref{eq:integral-1}, the analytical expression is too lengthy to be practical. One possible solution is to use the improved approximation for the error function that \cite{karagiannidis2007improved}
\begin{align}
  \erfc(x)\approx \frac{(1-e^{-Ax})e^{-x^2}}{B\sqrt{\pi}x},
\end{align}
where $A=1.98$ and $B=1.135$. Therefore,
\begin{align}
  \int_0^\infty dx \ x^ke^{-ax^2-bx}\erfc(cx)\approx\frac{1}{B\sqrt{\pi}}  \int_0^\infty dx \ x^{k-1}(1-e^{-Ax})e^{-(a+1)x^2-bx},
\end{align}
which is a Gaussian integral and clearly would be integrated in a closed-form.
\section{Quantum kicked rotor on resonance}
\label{sec:app3}
From the eigenequation $ U|\phi\rangle = e^{-i\epsilon}|\phi\rangle $ and the expression for Bloch states given in Eq. \eqref{eq:bloch}, we have

\begin{align}
  e^{-i\epsilon}\phi_{s,\alpha}&=\langle s|U|\phi\rangle=\sum_{n'} U_{s,n'}\phi_{n'}
  =\sum_{s'=1}^N\sum_{l'=-\infty}^{\infty} U_{s,s'+Nl'}\phi_{s'+Nl'} \nonumber\\
  &= \sum_{s'=1}^N\sum_{l'=-\infty}^{\infty}U_{s,s'+Nl}e^{-il'\alpha}\phi_{s',\alpha}\equiv \sum_{s'=1}^N (U_\alpha)_{s,s'}\phi_{s',\alpha}.
\end{align}
It reveals that $\phi_{s,\alpha}$ is the eigenstate of a reduced $N\times N$ unitary matrix $U_\alpha$ with
\begin{align}
  (U_\alpha)_{s,s'} \equiv \langle s|U_\alpha|s'\rangle =\sum_{l=-\infty}^\infty U_{s,s'+Nl} e^{-il\alpha}
  =e^{-is^2\thbar}\frac{1}{2\pi}\int_0^{2\pi}d\alpha' e^{-i(s-s')\alpha'}e^{-iK\cos(\alpha')/\thbar}\sum_{l=-\infty}^\infty e^{iNl\alpha'-il\alpha},
\end{align}
by making use of the Poisson summation formula $\frac{1}{2\pi}\sum_{l=-\infty}^\infty e^{ilx}=\sum_{j=-\infty}^\infty \delta(x-2\pi j)$, the matrix elements of ${U}_\alpha$ are 
\begin{align}
  \label{eq:reduce-floquet}
(U_\alpha)_{s,s'}&=\exp(\frac{-is^2\thbar}{2})\int_0^{2\pi}d\alpha' e^{-i(s-s')\alpha'}e^{-iK\cos(\alpha')/\thbar}\sum_j\delta(N\alpha'-\alpha-2\pi j) \nonumber\\
&=\frac{1}{N}\exp(\frac{-is^2\thbar}{2}-\frac{i(s-s')\alpha}{N}) \sum_{j=0}^{N-1} \exp(\frac{-i2\pi(s-s')j}{N}-\frac{iK\cos\frac{\alpha+2\pi j}{N}}{\thbar})\nonumber\\
&=\frac{1}{N}\exp(\frac{-is^2\thbar}{2}) \sum_{j=0}^{N-1} \exp(-{i(s-s')}\frac{\alpha+2\pi j}{N}-\frac{iK\cos\frac{\alpha+2\pi j}{N}}{\thbar}).
\end{align}

An important conclusion from Eq. \eqref{eq:reduce-floquet} is that $U_\alpha$ depends only on $N$ discrete points spaced in the interval $[0, 2\pi)$ in $\theta$ space with $\theta_j=(\alpha+2\pi j)/N$, with $\alpha/N$ being the initial point of each set of these points, and map onto each other under the operation of $U_\alpha$ for a fixed $\alpha$.  In the language of solid-state physics, it defines the dynamics of a particle in a finite one-dimensional lattice of size $N$, and different values of $\alpha$ correspond to different boundary condition. $N$ values of quasi-energy $\epsilon$ can be obtained from the diagonalization of $U_\alpha$, and by varying $\alpha$ in $[0, 2\pi)$ we have $N$ Floquet bands. In the following, we choose $\alpha=0$ the periodic boundary condition, resulting the parity in $U_0 \equiv U_\alpha\ (\alpha=0)$. Correspondingly, the quasi-energy eigenstates in phase representation have to be either even or odd $\phi(\theta)=\langle \theta|\phi\rangle =\pm\phi(-\theta)$, which in the momentum representation has the form $\phi_n =\pm \phi_{-n}$, and in the reduced momentum space $ \phi_s =\pm \phi_{-s}=\pm \phi_{N-s}$, one can restrict the study of $U_0$ to the invariant subspace of odd states, the matrix elements are
\begin{align}
  \langle s|\mathcal{P}_-U_0 \mathcal{P}_-|s'\rangle &=\langle s,-|U_0|s',-\rangle 
  = \frac{1}{N}\exp(\frac{-is^2\thbar}{2}) \sum_{j=0}^{N-1}\exp\big(\frac{-iK\cos\frac{2\pi j}{N}}{\thbar}\big)
\big(e^{-i2\pi (s-s')j/N}-e^{-i2\pi (s+s')j/N}+h.c.\big)/2\nonumber\\
&=\frac{1}{N}\exp(\frac{-is^2\thbar}{2}) \sum_{j=0}^{N-1}\exp\big(\frac{-iK\cos\frac{2\pi j}{N}}{\thbar}\big)\left[\cos\frac{2\pi(s-s')j}{N}-\cos\frac{2\pi(s+s')j}{N}\right],
\end{align}
where the projection operators $\mathcal{P}_\pm|s\rangle=|s,\pm \rangle =(|s\rangle \pm |N-s\rangle)/\sqrt{2}$, and $1\le s,s' \le N_1-1$ for $N=2N_1$.  Matrix elements for the even parity can be written similarly, noting that the basis states $|N/2,+\rangle$ and $|N,+\rangle = |0,+\rangle$ are not normalized.
\end{widetext}

\section{Matrix elements in circular Fock basis for cubic coupling terms}
  \label{sec:app4}
  Matrix elements in the circular basis of the cubic coupling terms are (for detailed derivation, see Ref. \cite{yan2024chaos})
  \begin{align}
      \label{eq:coeff-cubic}
    \langle n',l'|\hat{q}_\pm^3|n,  l\rangle=\hbar^{3/2}\delta_{l',l\pm3}\sum_{m} k_m^\pm (n,l)\delta_{n',n+m},
  \end{align}
  where $m\in \mathcal{M}_1=\{\pm 1, \pm3\}$, $k_m^+(n,l)=k_m^-(n,-l)$, and 
  \begin{align}
  \begin{aligned}
      \label{eq:no-phase}
      &k_{-1}^+=3\sqrt{(n-l)(n-l-2)(n+l+2)/8}, \\
      &k_{1}^+=3\sqrt{(n-l)(n+l+2)(n+l+4)/8},  \\
      &k_{-3}^+=\sqrt{(n-l)(n-l-2)(n-l-4)/8},  \\
      &k_{3}^+=\sqrt{(n+l+2)(n+l+4)(n+l+6)/8}.
  \end{aligned}
\end{align}

The cubic terms couple states with $\Delta l=\pm 3$, therefore the coupling takes place in three decoupled sets of basis states: the singlet $\{a\}$ $l\in \{\dots -6, -3 ,0, 3, 6\dots\}$, two doublets with the same eigenspectra $\{b\}$ $l\in \{\dots -5, -2 ,1, 4\dots\}$ and $\{c\}$ $l\in \{\dots -4, -1 ,2, 5\dots\}$. This property of eigenspectra can also be verified from the $C_{3v}$ symmetry of the classical Hamiltonian, that they must belong to the irreducible representations of the point symmetry group $C_{3v}$: the subspaces of two doublets are of $E$ symmetry and degenerate, and the singlet is a combination of ($A_1$, $A_2$) symmetry. 
\end{appendix}

\bibliography{ratioMixed.bib}
\end{document}